\documentclass{jfmmod}
\usepackage{graphicx}
\usepackage{epstopdf, epsfig}
\usepackage{amsmath}
\usepackage{amsfonts}
\usepackage{amssymb}
\usepackage{float}

\usepackage{hyperref}

\usepackage{amsmath}
\usepackage{amsfonts}
\usepackage{amssymb} 
\usepackage{graphicx}

\usepackage[normalem]{ulem}

\usepackage{pdfpages}

\definecolor{colortodo}{RGB}{250,0,0}
\definecolor{colorblack}{RGB}{0,0,0}

\def\SL{{S_\textrm{L}}}
\def\ST{{S_\textrm{T}}}

\def\wRath{{\omega_{\textrm{R},\alpha}^\textrm{th}}}
\def\wRL{{\omega_\textrm{R,L}}}
\def\LambdaL{{\Lambda_\textrm{L}}}

\def\wRT{{\omega_\textrm{R,T}}}
\def\LambdaT{{\Lambda_\textrm{T}}}

\def\wAL{{\omega_\textrm{A,L}}}
\def\wAT{{\omega_\textrm{A,T}}}
\def\wD{{\omega_\textrm{D}}}
\def\wR{{\omega_\textrm{R}}}
\def\wA{{\omega_\textrm{A}}}

\def\woL{{{\omega_\textrm{0,L}}}}
\def\woT{{{\omega_\textrm{0,T}}}}
\def\CQ{{C_\textrm{Q}}}
\def\CQL{{C_\textrm{Q,L}}}
\def\CQT{{C_\textrm{Q,T}}}
\def\CA{{C_\textrm{A}}}
\def\CAL{{C_\textrm{A,L}}}
\def\CAT{{C_\textrm{A,T}}}

\def\CG{{C_\textrm{G}}}

\def\CGL{{C_\textrm{G,L}}}
\def\CGT{{C_\textrm{G,T}}}

\def\CPL{{C_\textrm{P,L}}}

\def\KA{{K_\textrm{A}}}
\def\KQ{{K_\textrm{Q}}}
\def\GA{{G_\textrm{A}}}
\def\GQ{{G_\textrm{Q}}}

\def\Ek{{E_\textrm{K}}}
\def\Em{{E_\textrm{m}}}
\begin{document}

%%%% Article title to be placed here
%\title{Wave spectroscopy in a driven granular material}
%
%
%\author{Michael Berhanu}
%\affiliation{Universit\'e Paris Cit\'e, CNRS, MSC, UMR 7057, F-75013, Paris, France}
%\author{Simon Merminod}
%\affiliation{Department of Molecular and Cellular Biology, Harvard University, Cambridge, MA~02138, USA}
%\author{Gustavo Castillo}
%\affiliation{Instituto de Ciencias de la Ingenier\'ia, Universidad O'Higgins, 2841959 Rancagua, Chile}
%\author{Eric Falcon}
%\affiliation{Universit\'e Paris Cit\'e, CNRS, MSC, UMR 7057, F-75013, Paris, France}

\shorttitle{Wave spectroscopy in a driven granular material}
\shortauthor{Michael Berhanu et al.}

\title{Wave spectroscopy in a driven granular material}

\author{Michael Berhanu\aff{1}
  \corresp{\email{michael.berhanu@univ-paris-diderot.fr}},
  Simon Merminod\aff{2},  Gustavo Castilloh\aff{3},
 \and Eric Falcon\aff{1}}

\affiliation{\aff{1}Universit\'e Paris Cit\'e, CNRS, MSC, UMR 7057, F-75013, Paris, France
\aff{2}Department of Molecular and Cellular Biology, Harvard University, Cambridge, MA~02138, USA
\aff{3}Instituto de Ciencias de la Ingenier\'ia, Universidad O'Higgins, 2841959 Rancagua, Chile}

%%%%% Subject entries to be placed here %%%%
%\subject{Physics, Granular media, wave physics}
%
%%%%% Keyword entries to be placed here %%%%
%\keywords{ Driven granular media, Dispersion relations, Phonons, Elastic coefficients}

%%%% Insert corresponding author and its email address}
%\corres{Michael Berhanu\\
%\email{michael.berhanu@univ-paris-diderot.fr}

\maketitle

%%%% Abstract text to be placed here %%%%%%%%%%%%
\begin{abstract}
Driven granular media constitute model systems in out-of-equilibrium statistical physics. By assimilating the motions of granular particles to those of atoms, by analogy, one can obtain macroscopic equivalent of phase transitions. Here, we study fluid-like and crystal-like two-dimensional states in a driven granular material. In our experimental device, a tunable magnetic field induces and controls remote interactions between the granular particles. We use high-speed video recordings to analyse the velocity fluctuations of individual particles in stationary regime. Using statistical averaging, we find that the particles self-organize into collective excitations characterized by dispersion relations in the frequency-wavenumber space. These findings thus reveal that mechanical waves analogous to condensed matter phonons propagate in driven granular media. When the magnetic coupling is weak, the waves are longitudinal, as expected for a fluid-like phase. When the coupling is stronger, both longitudinal and transverse waves propagate, which is typically seen in solid-like phases. We model the dispersion relations using the spatial distribution of particles and their interaction potential. Finally, we infer the elastic parameters of the granular assembly from equivalent sound velocities, thus realizing the spectroscopy of a granular material.
\end{abstract}
%%%%%%%%%%%%%%%%%%%%%%%%%%%

%%%%%%%%%% Insert the texts which can accomdate on firstpage in the tag "fmtext" %%%%%

%\begin{fmtext}
%%
%\end{fmtext}

\section{Introduction}

Granular systems are defined as assemblies of separate, macroscopic and athermal particles~\cite{Andreotti2013}. Such systems occur in Nature under a variety of forms: from the sand of beaches and dunes on Earth to small astronomic bodies---interplanetary dust, meteoroids, and small asteroids---in space. But despite the ubiquity of granular systems, a unified description of the collective motions of their particles is still missing~\cite{Mujica2016}. 
A reason for this is the complexity due to energy dissipation through particle-particle and particle-substrate friction and dissipative collisions. Such energy dissipation also means that granular systems require continuous inputs of energy to sustain their dynamics. Granular systems thus constitute model systems to study matter out of the thermal equilibrium~\cite{Mujica2016}. 

As it can be observed in everyday life, granular matter can be considered as a solid, liquid or gas, depending on the particle density and the amount of energy injected into the system, for instance mechanically. In molecular systems, phase transitions result from the competition between thermal agitation and the interaction between the particles. In granular systems, the macroscopic counterpart to phase transitions~\cite{Jaeger1996,Aranson2006,Andreotti2013} can be obtained by varying the number of particles~\cite{Falcon1999a,Noirhomme2018,Noirhomme2021}, inducing interactions between particles~\cite{Aranson2000,Blair2003} or  tuning the confinement~\cite{Schmidt1997}. In particular, thin and vibrated layers of grains~\cite{Olafsen1998,Prevost2004,Clerc2008,Castillo2012,Mujica2016,Opsomer2019} are of fundamental interest because they allow using particle tracking techniques, yielding the dynamics of each grain individually and over durations far exceeding those accessible through molecular dynamics simulations. These studies bridge the gap between a description at the particle scale and a global analysis of collective motions.

By analogy with the hydrodynamics description of molecular systems~\cite{HansenMcdonaldBook}, we characterize a driven granular system in a non-equilibrium stationary state from the analysis of the particle velocity fluctuations in the Fourier space. We previously introduced a model experimental system~\cite{Merminod2014,Merminod2015,Castillo2020} to study the competition between mechanical agitation and remote interactions in a quasi-two-dimensional quasi-2D assembly of magnetized, millimeter-diameter spheres. For a given cell filling and shaking strength, we observed a granular fluid-like phase or a solid-like crystal phase depending on the strength of a tunable dipolar repulsion~\cite{Merminod2014}. Here, we show that the spontaneous collective excitations in this system can be described as a set of random propagative waves characterized by dispersion relations. Thus, we obtain a macroscopic equivalent of phonons, which correspond to the standard decomposition of atomic vibrations under the form  of normal modes in condensed matter~\cite{Ashcroftbook}. Phonons propagate in both crystalline lattices and amorphous matter. In our experiment, we take advantage of the propagation of these waves to extract the elastic coefficients characterizing the mechanical behaviour of the granular material. Therefore, our study constitutes a sound spectroscopy, which gives access to the properties of the material using a non-intrusive method, \textit{i.e.}, without submitting it to a mechanical perturbation such as a shear. A similar approach using particle tracking has been recently implemented in active matter for a system of self-propelled colloidal rollers undergoing collective motion and which validated the hydrodynamics description of active matter~\cite{Geyer2018}. 

Collective excitations have also been studied in dusty plasmas, also called complex plasmas, which are constituted of macroscopic, charged particles in levitation and interacting through a screened Coulombian potential~\cite{Fortov2005,Morfill2009}. In 2D complex plasma crystals, phonon spectra of longitudinal and transverse waves are generated by random particle motions~\cite{Nunomura2002,Couedel2009,Couedel2019}. More recently, numerical and theoretical works~\cite{Golden2008,Golden2010,Khrapak2018,Mistryukova2019} investigated the collective excitations of 2D assemblies of interacting dipoles ; however, these results are yet to be experimentally validated. For dissipative granular gases, a few numerical simulation works~\cite{Vollmayr2011,Brito2013} computed space-time spectra in dissipative granular gases, and showed that velocity fluctuations organize into longitudinal waves. These works extracted effective macroscopic coefficients ---viscosity and diffusivity---  which they compared with the theoretical predictions from the fluctuating hydrodynamics theory of driven granular media~\cite{Sela1998,vanNoijeErnst1999,Goldhirsch2003,Gradenigotheo2011,Puglisibook}. However, to our knowledge, an experimental validation of these theoretical predictions is still missing. 
In colloidal crystals, the mechanical constants of the medium can be extracted from the fluctuations of particle displacements in response to thermal agitation~\cite{Sengupta2000,Zahn2003,Grunberg2004,Gasser2010}, but this requires assuming thermal equilibrium. A similar approach has been applied to granular particles with electrostatic interactions~\cite{CoupierPhD2006}. In this article, the wave spectroscopy that we carry out relies on the analysis of velocity fluctuations in the spatial and temporal Fourier spaces; it thus necessitates high-speed imaging with a frame rate of about $10^3$ Hz. This method does not require thermal equilibrium and is thus advantageous for the out-of-equilibrium, driven granular material studied here.

\section{From granular gas to hexagonal crystal}
\begin{figure}
\centering
\includegraphics[width=.7\linewidth]{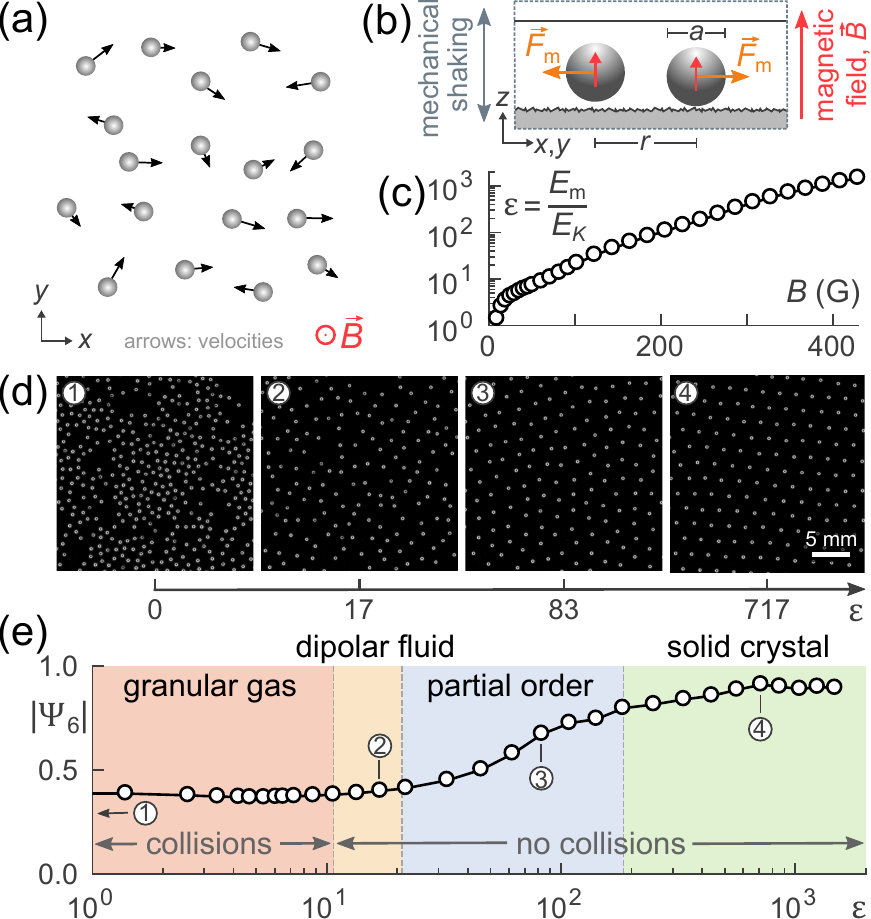}
\caption{{Overview of our experimental system.
(a)~Top view schematics.
Millimetre-sized particles follow Brownian-like motions in quasi-two-dimensions (black arrows) in the absence of magnetic field, and interact via remote interactions when immersed in a transverse, operator-tuned magnetic field, $\mathbf{B}$.
(b)~Side view schematics of a region of the experimental cell. As the cell is vertically shaken (double grey arrow), the chrome steel beads rebound on the rough bottom surface in random directions and velocities. When the magnetic field immersing the cell is turned on (long red arrow), the beads become induced magnetic dipoles (short red arrows) and repel each other (orange arrows).
The bead diameter is $a=1$~mm, the vertical gap is roughly $1.5a$, and the root mean square roughness of the bottom surface is 20~$\mu$m.
(c)~Dimensionless parameter $\varepsilon$, which quantifies the competition between magnetic interactions and bead kinetic energy, increases with the amplitude of the magnetic field $B = ||\mathbf{B}||$.
(d)~Top-view photographs show increasing order in the bead assembly as $\varepsilon$ increases and dipolar forces strengthen.
(e)~The sixfold bond orientational parameter, $|\Psi_6$|, monotonically increases as a function of $\varepsilon$. 
We define four distinct regimes: ``granular gas'' (bead collisions occur, $\varepsilon <10.7$), ``dipolar fluid'' (no collisions and low $|\Psi_6$|, $10.7<\varepsilon<21.9$), ``partially ordered phase'' abbreviated as ``partial order'' (no collisions and steeply increasing $|\Psi_6$|, $21.9<\varepsilon<185$), ``solid crystal'' (no collisions and plateauing $|\Psi_6$|, $\varepsilon >185$). Circled numbers correspond to the conditions of the photographs in panel~(d).
\\ } 
}
\label{FigSetup}
\end{figure}

Our experimental system consists of a quasi-monolayer of millimetre-diameter granular spheres that are vibrated on a rough plate and can be magnetized through an external magnetic field (figure ~\ref{FigSetup}a-b). We used $N_0=2000$ spherical particles of diameter $a=1\,$mm and made of chromed steel, which is a soft-ferromagnetic material. These particles form a monolayer inside a horizontal, square cell of internal dimensions roughly 9~cm $\times$ 9~cm $\times$ 1.5~mm. By vibrating  the cell vertically with a frequency of 300 Hz and an acceleration of $1.6g$, where $g$ is the gravitational acceleration, the particles acquire kinetic energy and their motions projected onto the horizontal plane are quasi-Brownian~\cite{Olafsen1998,Losert1999,Reis2007,Merminod2014}. 
Using a high-speed camera and tracking algorithms (see Materials and Methods) we compute the positions and velocities of individual particles within a region of interest of area $50.36 \times 50.36~\text{mm}^2$ around the cell center. 

Furthermore, we can turn on an external magnetic field, $\mathbf{B}$, which is vertical---\textit{i.e.}, perpendicular to the bottom of the cell---so that each particle becomes an induced magnetic dipole (figure ~\ref{FigSetup}b~). Thus, two particles located in a same horizontal plane and whose centres are separated by a distance $r_{i,j}$ interact through the repulsive potential~\cite{Jackson1999}
\begin{equation}
U_{i,j}=\frac{4\pi}{\mu_0}\,B^2 \frac{(a/2)^6 }{r_ {i,j}^3}\, ,
\label{Uij}
\end{equation}
where $\mu_0$ is the vacuum permeability constant and $B = ||\mathbf{B}||$. %Note that the induced magnetization between magnetized particles is negligible~\cite{Opsomer2019}.

Throughout this article, we use a single dimensionless parameter, $\varepsilon$, to quantify the competition between kinetic energy and magnetic repulsion in the particle assembly. 
Specifically, we define $\varepsilon\equiv\Em/\Ek$, where $\Em$ is the mean magnetic energy per particle (Eq.~\ref{Em}) and $\Ek$ is the mean kinetic energy per particle (Eq.~\ref{Ek}).
Parameter $\varepsilon$ is analogous to an inverse temperature and constitutes our control parameter. We carried out 31 experiments with $B$ between $[0,430]$~G, corresponding to values of $\varepsilon$ between $[0,1500]$ (figure ~\ref{FigSetup} c)~).

As we previously showed~\cite{Merminod2014,Castillo2020}, increasing $\varepsilon$---through $B$---induces a gas-like to solid-like transition in the granular quasi-monolayer (figure ~\ref{FigSetup}d). More specifically, when $\varepsilon=0$, the combination of homogeneous mechanical forcing and inherently dissipative collisions results in a granular gas state~\cite{Olafsen1998,Losert1999,Reis2007,Puglisi2012,Mujica2016}. From $\varepsilon \sim 10$, we do not detect collisions anymore and a fluid of interacting dipoles is formed~\cite{Merminod2014,Castillo2020}. At the highest values of $\varepsilon$, particles self-organize into a hexagonal, crystal-like structure. Notably, a study with a similar experiment as our own~\cite{Schockmel2013} showed that this crystallization follows the Kosterlitz-Thouless-Halperin-Nelson-Young (KTHNY) scenario~\cite{Strandburg1981,Grunberg2007,Gasser2010}.

We define four distinct regimes of the granular assembly based on the degree of orientational order and the collision rate in the system.
As a measure orientational order, we use the time- and ensemble-averaged sixfold bond-orientational order parameter, or global hexagonal order parameter, $|\Psi_6|$  (Eq.~\ref{Psi6b}). $|\Psi_6|$ equals~1 for a perfectly hexagonal lattice and is smaller otherwise. We find that $|\Psi_6|$ increases from roughly 0.4 to 0.9 across the full range of $\varepsilon$ (figure ~\ref{FigSetup}e).
First, we define the ``granular gas'' state for $0<\varepsilon<10.7$, where $|\Psi_6|$ is the smallest, collisions occur---although at a decreasing rate as $\varepsilon$ increases---and the particle assembly behaves as a dissipative granular gas~\cite{Castillo2020} (Movies S1 and S5). 
Second, for $10.7<\varepsilon<21.9$, the ``dipolar fluid'' regime consists in a disordered fluid of strongly coupled dipoles, for which $|\Psi_6|$ is still near 0.4 but there are no more collisions (Movies S2 and S6). Magnetic energy then significantly dominates kinetic energy, and the system behaves similarly to dusty plasmas~\cite{Donko2008,Morfill2009}.
Third, in the ``partial order'' regime, for $21.9<\varepsilon<185$, the hexagonal order parameter $|\Psi_6|$ increases from roughly 0.4 to 0.8, with the maximum variation in $|\Psi_6|$ occurring near $\varepsilon = 62$ (Movies S3 and S7).
Finally, for $\varepsilon > 185$, we define the ``solid crystal'' state as hexagonal order is high, with $|\Psi_6|>0.8$ (Movies S4 and S8). There, while particles do not exchange neighbours, they vibrate around their local magnetic energy minimum, which together define the hexagonal lattice, due to the mechanical shaking. Note that defects are present in this crystal-like structure, but they likely do not significantly impact the dynamical properties which are the focus of this article.
See the electronic supplementary material Sections~5 and~6 for a more detailed characterization of these four regimes using the pair correlation function and the mean-square displacements.

Collective excitations correspond to a coherent behaviour of these vibrations evidenced using a representation in the Fourier space. Such excitations have been studied theoretically and numerically for 2D systems of interacting dipoles~\cite{Golden2008,Golden2010,Khrapak2018,Mistryukova2019}. In particular, Khrapak \textit{et al.}~\cite{Khrapak2018} studied in details the dynamics and thermodynamics of such systems, depending on a parameter equivalent to our $\varepsilon$, and for both strongly coupled dipolar fluid and solid phases; their work provides a theoretical guide for our experimental study.

\section{Experimental spectra of velocity fluctuations}

\begin{figure*}[t]
\begin{center}
\includegraphics[width=1\linewidth]{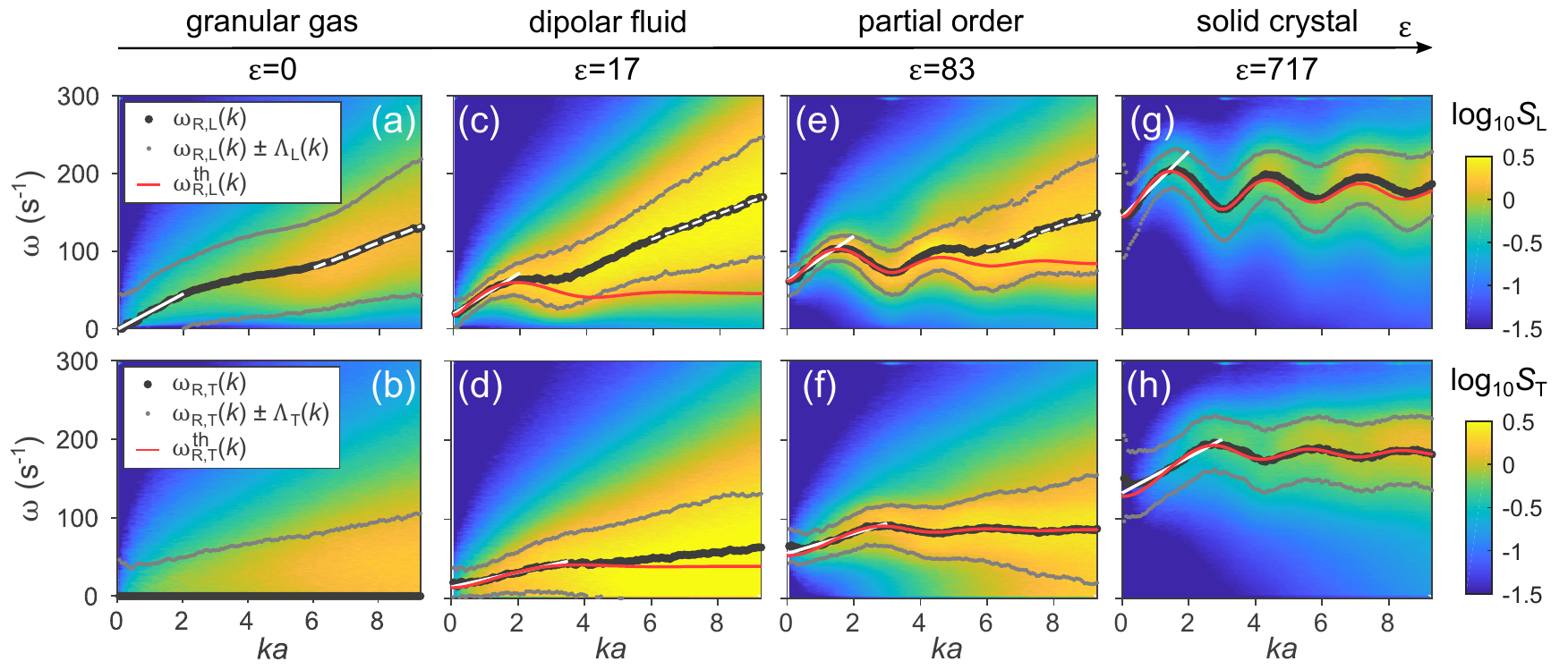}
\caption{
Two-dimensional wave propagation.
(a,c,e,g)~Space-time spectra of longitudinal excitations, ${\SL}(k,\omega)$, in each of the four regimes.
(b,d,f,h)~Space-time spectra of transverse excitations, ${\ST}(k,\omega)$. 
(a--h) In each panel, we obtain an experimental dispersion relation, ${\wRL}(k)$ (respectively, ${\wRT}(k)$, black bullets), and its width, ${\LambdaL}(k)$  (resp., ${\LambdaT}(k)$, grey bullets) by fitting a double-Lorentzian function (Eq.~\ref{doubleLorentz}) to the space-time spectra, $\SL$ (resp., $\ST$).
The solid white lines at small $k$ show linear fits of slope $\CGL$ $ \equiv \partial \wRL/\partial k |_{k \rightarrow 0}$ (resp. $\CGT \equiv \partial \wRT/\partial k |_{k\rightarrow 0}$) 
The dashed white lines at large $k$ show linear fits of slope $\CPL$ $ \equiv \partial \wRL/\partial k$.
The red curves show fits of the experimental dispersion relations according to Eq.~\ref{Equafit}, with a single adjustable parameter for each. Eq.~\ref{Equafit} is derived from the theoretical dispersion relation for a 2D array of point dipoles interacting with each other~\cite{Khrapak2018}. (a,b) show that the granular gas has properties typical of a fluid: propagation of acoustic longitudinal waves~(a) and absence of transverse waves~(b). (c--h) show that dipolar forces between the particles enable the propagation of both longitudinal and transverse waves.}
\label{figSpectra}
\end{center}
\end{figure*}

We now investigate the collective excitations propagating in the granular medium in stationary regime. For that, as in studies of the hydrodynamic limit of molecular liquids~\cite{HansenMcdonaldBook} and driven granular media~\cite{Gradenigotheo2011,Brito2013}, we start by computing a current density ${\mathbf{j}}$, which is the coarse grained velocity field associated with the motions of individual particles, and whose space Fourier transform is ${\mathbf{\tilde{j}}} (\mathbf{k},t)=\sum_i^N \, \mathbf{v}_i (t)\, \mathrm{e}^{\imath \, \mathbf{k} \cdot  \mathbf{r}_i (t)}$, with $\mathbf{k}$ the 2D wave number, $\mathbf{r}_i$ the position of particle $i$, and $\mathbf{v}_i$ the velocity of particle $i$. We also assume isotropy, allowing us to consider the dynamics of the longitudinal currents (\textit{i.e.}, verifying $\mathbf{k} \times \mathbf{r}_i = 0$) and  transverse currents ($\mathbf{k} \cdot \mathbf{r}_i = 0$) decoupled. Then, we identify the collective modes and characterize their dynamics by computing the space-time spectra  (Eq.~\ref{spectraSvl}) of longitudinal and transverse excitations, respectively $\SL (k, \omega)$ and $\ST (k, \omega)$, where $k = ||\mathbf{k}||$. The wavenumber $k$ and the angular frequency $\omega$ are the natural variables to study wave propagation.

Space-time spectra representative of the the four regimes we identified in figure ~\ref{FigSetup}e---granular gas, dipolar fluid, partial order, and solid crystal---are shown in figure ~\ref{figSpectra}. Under each condition, the ensemble of the energy maxima across the range of $ka$ defines a curve called the dispersion relation. By fitting each space-time spectra with a double Lorentzian function (Eq.~\ref{doubleLorentz}), we extract experimental dispersion relations, $\wRL (k)$ and $\wRT (k)$, and their widths, $\LambdaL (k)$ and $\LambdaT (k)$, respectively. 

In the absence of magnetic field ($\varepsilon=0$) and for the longitudinal spectrum, the energy is distributed around a non-vanishing dispersion relation (figure ~\ref{figSpectra}a). For $ka < 2$, we find that longitudinal waves have an acoustic behaviour, \textit{i.e.}, $\wRL (k)=\CGL\,k$, where $\CGL$ is a longitudinal sound velocity. Thus we show experimentally, that the velocity fluctuations in a driven granular medium without remote interactions correspond to the propagation of sound waves similar to those in a molecular gas. {Such observation is predicted by the hydrodynamic theory for homogeneously driven granular gas with dissipative collisions~\cite{vanNoijeErnst1999,Gradenigotheo2011} and has been also reported in event-driven numerical simulations~\cite{Vollmayr2011,Brito2013}.} For $k a > 6$, the dispersion relation is fitted by $\wRL (k)=\CPL\,k+ \Omega$, where $\CPL$ is a sound velocity and $\Omega$ is a constant, denoting a dispersive propagation. In contrast, for the transverse spectrum, the energy maximum is at $\omega=0$ at all $ka$, indicating the absence of transverse waves, as expected in a fluid phase (figure ~\ref{figSpectra}b).

As the magnetic field is increased, the acoustic behaviour at low $k$ is lost, \textit{i.e.}, $\wRL (k)=\CGL\,k + \omega_\textrm{0,L}$, with $\omega_\textrm{0,L}$ a cut-off frequency and $\CGL$ the fitted group velocity. Moreover, the dispersion relation becomes non-monotonic, thus showing regions of negative group velocity and even oscillations when interactions are the strongest (figures ~\ref{figSpectra}c--h). In dispersion relations, local minima due to regions of a negative group velocity are called rotons. While rotons have recently been observed in a quantum gas~\cite{Chomaz2018}, our results show that quantum effects are not necessary ingredients for rotons---a finding also in agreement with studies attributing the roton minimum to position correlations~\cite{Kalman2010}. 

For longitudinal waves and moderate $\varepsilon$ \ (figure ~\ref{figSpectra}c,e), we observe again a linear dispersion relation for $ka > 6$, namely $\wRL (k)=\CPL\,k + \Omega$, where $\CPL$ is a sound velocity and $\Omega$ is a constant, which is consistent with numerical simulations of interacting, thermally agitated dipoles~\cite{Golden2010,Khrapak2018}. This branch corresponds to lengthscales smaller than the typical distance between particles and we attribute it to single-particle excitations~\cite{Golden2010,Khrapak2018}. In contrast, in the solid phase, the dispersion relation is non-monotonic even for large $ka$ (figure ~\ref{figSpectra}g).

Importantly, the presence of a magnetic field enables transverse waves to propagate as well. We observe indeed dispersion relations for transverse waves, $\wRT (k)$, in all regimes but the granular gas {(figure ~\ref{figSpectra}d,f,h)}. This shows that, even with moderate magnetic interactions as in the dipolar fluid regime, the system becomes rigid enough to transmit transverse displacements as commonly observed in solids and in agreement with numerical simulations of 2D arrays of interacting dipoles~\cite{Golden2010,Khrapak2018}. Again, we report a non-acoustic behaviour of transverse waves at low $k$: $\wRT (k)=\CGT\,k + \omega_\textrm{0,T}$, due to the presence of a cut-off frequency $\omega_\textrm{0,T}$. 

We model our experimental dispersion relations based on a theory for wave propagation in dipole arrays. The theoretical dispersion relations for a 2D array of point dipoles within the harmonic potential approximation, $\omega_\textrm{Q,L}(k)$ and $\omega_\textrm{Q,T}(k)$, can be analytically computed using the quasilocalized charge approximation (QLCA) theory~\cite{Golden2008,Golden2010,Khrapak2018} (see Materials and Methods Eqs.~\ref{eqwQL} and~\ref{eqwQT}). %and electronic supplementary material Section 3 for more details)
% \red{~(the formulas of $\omega_\textrm{Q,L}(k)$ and $\omega_\textrm{Q,T}(k)$  are given in electronic supplementary material Section 3)}. 
The QLCA theory involves a characteristic dipole oscillation frequency,
\begin{equation}
    \wD=B\,\phi^{5/4}\,\sqrt{\dfrac{4\pi\,a}{\mu_0\,m}}{,}
    \label{omegaD_main}
\end{equation}
with $\phi$ the area fraction of dipoles. Note that $\wD$ increases linearly with the applied magnetic field $B$. Notably, in our experiments, $\wRL (k)$ and $\wRT (k)$ do not tend towards zero when $k$ does so. Instead, cut-off frequencies at $k=0$, namely $\woL$ for longitudinal waves and $\woT$ for transverse waves, prevent acoustic wave propagation at large scales. Therefore, we use the following modified version of the QLCA theory results to model our experimental dispersion relations:

\begin{equation}
    \wRath^2 (k)=\left[\dfrac{\omega_{\mathrm{A},\alpha}}{\wD}\,\omega_{\mathrm{Q},\alpha} (k)\right]^2+\omega_{\textrm{0},\alpha}^2\, ,
    \label{Equafit}
\end{equation}
where the subscripts $\alpha$ should be substituted by ``L'' for longitudinal waves, and ``T'' for transverse waves; and $\omega_{\mathrm{A},\alpha}$ are the best-fit parameters that are our experimental estimates correcting the QLCA characteristic frequency, $\wD$.

These semi-empirical dispersion relations describe increasingly well our experimental dispersion relations as remote particle interactions strengthen and order increases in the particle lattice.
Indeed, in the solid crystal regime, our model matches closely both longitudinal and transverse dispersion relations (figure ~\ref{figSpectra}g,h), while in the partial order regime, $\wRT$ is better fitted than $\wRL$ (figure ~\ref{figSpectra}e,f), and neither $\wRT$ or $\wRL$ are well fitted in the dipolar fluid regime, where orientational order is very low (figure ~\ref{figSpectra}c,d).

\begin{figure*}[t]
\begin{center}
\includegraphics[width=1\linewidth]{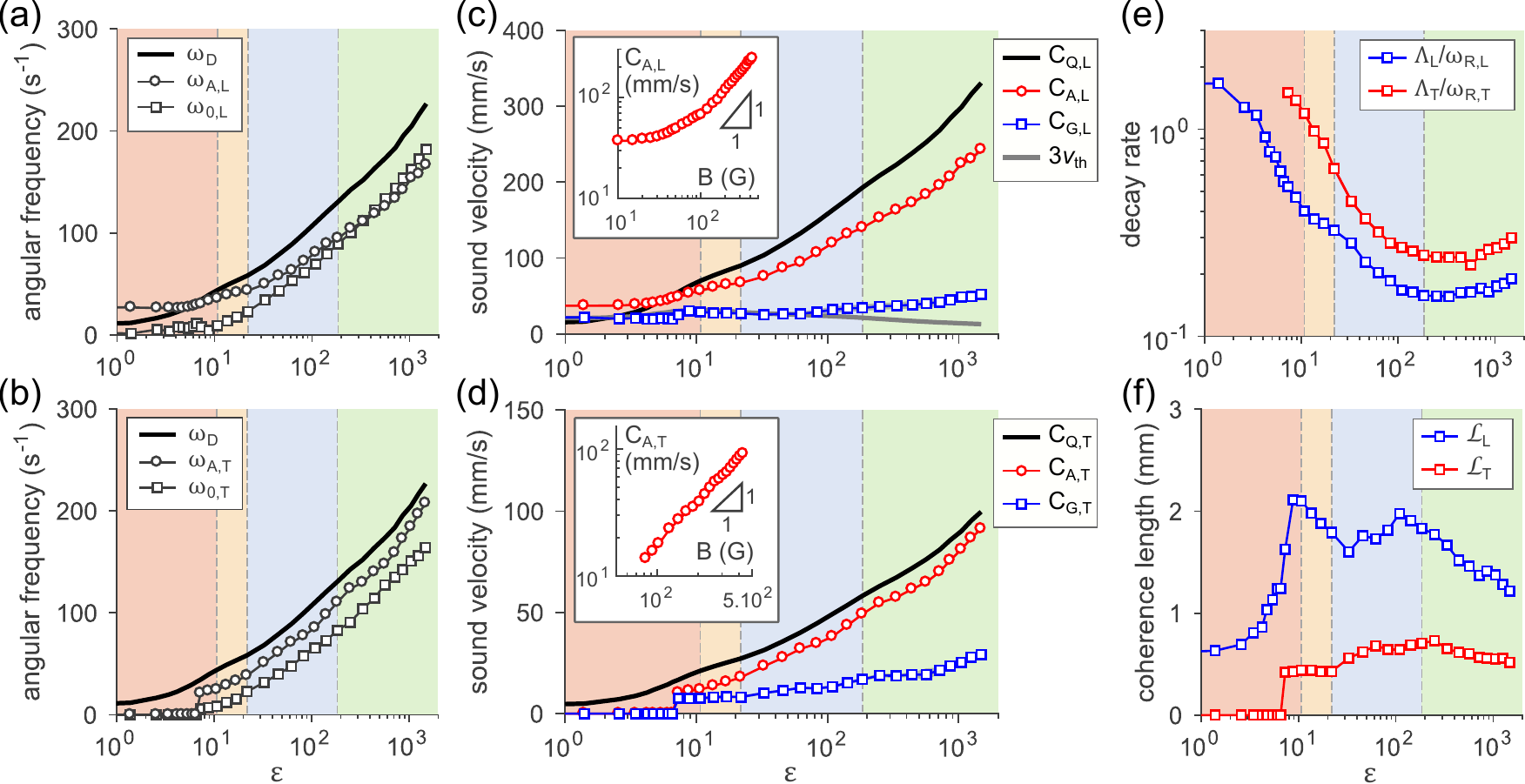}
\caption{
{Sound propagation.
(a,b)~Characteristic dipole frequency, $\wD$ {(}Eq.~\ref{omegaD_main}, solid curve), experimental cut-off frequency, $\omega_\textrm{0}$ (squares), and adjusted frequency $\wA$ (circles) obtained by fitting $\omega_\textrm{R}^\textrm{th}$ (Eq.~\ref{Equafit}) to each experimental dispersion relation, $\wR(k)$, for longitudinal~(A) and transverse~(B) waves.
(c,d)~Sound velocities as a function of $\varepsilon$ for longitudinal~(c) and transverse~(d) waves:
$\CQ$ (solid black curves) is obtained from the QLCA theory (see Material and Methods Eqs.~\ref{cQl} and \ref{cQt}); $\CA$ (red circles) is obtained from the same two equations, after $\wD$ was substituted by $\wA$;
$\CG$ (blue squares) is the group velocity, taken as the slope of the linear fits of the experimental dispersion relations at small $k$ (see solid white lines in figure ~\ref{figSpectra});
and $3v_\textrm{th}$ (solid grey curve) is three times the ``thermal'' velocity, that is, $3(E_\textrm{K}/m)^{1/2}$.
%is proportional to the root-mean-square velocity of individual particles.
Insets in (c,d) show that $\CA$ is roughly proportional to the magnetic field, $B$, when $B$ is strong.
(e)~The dimensionless decay rates at large scales, $\LambdaL/\wRL$ and $\LambdaT/\wRT$ (averaged within $0<ka<1.68$ for each $\varepsilon$), show that waves are strongly damped at all $\varepsilon$.
(f)~The coherence lengths, ${\mathcal{L}_{\mathrm{L}}}=\CGL/\LambdaL$ and ${\mathcal{L}_{\mathrm{T}}}=\CGT/\LambdaT$, reveal that waves propagate only on distances of the order of the particle diameter, $a=1\,$mm, at all $\varepsilon$.
}
}
\label{cplot}
\end{center}
\end{figure*}

\section{Sound velocities}

In this section, we discuss the relevance of our semi-empirical model for the experimental dispersion relations, and we investigate their main properties---namely sound velocities, excitation lifetimes, and excitation coherence lengths.

We start by comparing the characteristic frequency from the QLCA theory, $\wD$, with our experimental estimates $\wAL$ and $\wAT$ in figures ~\ref{cplot}a,b. In the granular gas regime, the comparison is not relevant because the dipolar interactions between the particles are negligible. However, as magnetic interaction increase between particles, the fitted frequencies become nearly proportional to $\wD$, with $\wAL \approx 0.75 \, \wD$ and $\wAT \approx 0.9 \, \wD$ and thus close to the value $\wD$ predicted by the QLCA theory.

The presence of cut-off frequencies in our system is not trivial and calls for an in-depth discussion. We find that the frequencies $\woL$ and $\woT$ become finite from near the onset of the dipolar fluid regime and grow proportionally to $\wD$ (figures ~\ref{cplot}a,b). Usually, such cut-offs are associated with an absence of  translational invariance, which is typical of finite size systems. Yet, here they vanish in the granular gas regime before they increase proportionally to $\wD$. Thus, these cut-off frequencies seem more likely to result from the dipolar interactions than from a geometrical constraint such as in a wave-guide. Particle vertical oscillations---unfortunately inaccessible with our 2D particle tracking---could occur due to the magnetic forces, influence in-plane wave propagation, and result in cut-off frequencies. Indeed, as long as the vertical gap in the experimental cell is larger than the particle diameter, $a$, the configuration of minimum magnetic potential energy for a local group of particles is where particles are in contact alternatively with the bottom surface and the lid~\cite{Merminod2015,MerminodPhd}, leading to vertical zigzag structures~\cite{Opsomer2019} (see electronic supplementary material Section 3).
We find the characteristic frequency associated with this energy minimum to be roughly 80\%--90\% of $\wD$. We show with a simple phenomenological model, that this coupling of horizontal motions with localized vertical oscillations results in dispersion relations with an acoustic branch for in-phase oscillations and an optical branch for antiphase oscillations (see electronic supplementary material Section 4). The acoustic branch of the dispersion relation intersects the origin, whereas for the optical branch has a cut-off frequency. In an analogous geometrical configuration, although using a repulsive electrostatic potential, acoustic and optical branches of longitudinal and transverse oscillations have also been obtained theoretically~\cite{Dessup2015}.
However, the extension of this work to 3D is challenging even in the crystal regime, because the hexagonal order is incompatible with the up-down alternation and leads to geometrical frustration~\cite{Han2008}. Experiments on quasi-2D dusty plasmas have demonstrated that out-of-plane waves create an optical branch of excitations~\cite{Couedel2009}, which for high forcing can be coupled with the longitudinal and transverses branches by mode coupling instability and induce crystal melting~\cite{Couedel2010,Couedel2019}.
Similarly,  for a 2D system of electrons trapped at the surface of liquid helium, the coupling of the longitudinal and transverse waves with the capillary waves modifies the dispersion relation of the optical phonons at low $k$~\cite{Grimes1979,Fischer1979,Marty1980,Gallet1982}.
The obtained dispersion relation in $\omega^2$ corresponds to the sum of the squares of the characteristic capillary frequency and of the 2D dispersion relation, as here in Eq.~\ref{Equafit}.
In our experiments, mechanical shaking appears to preferentially force optical wave modes, \textit{i.e.}, the horizontal and vertical oscillations are in antiphase for both the longitudinal and transverse excitations, which explains the cut-off frequencies.
An extension of the theory by Khrapak \textit{et al.}~\cite{Khrapak2018} to 3D would thus be useful to better understand wave propagation in confined layers.

We extract three types of sound velocities from the experimental dispersion relations to characterize the propagation of the waves (Figs.~\ref{cplot}c,d).
Firstly, the linear fits of the dispersion relations at small $k$ yield sound group velocities $\CGL$ and $\CGT$, as explained in the previous section.
Secondly, we estimate theoretical sound velocities using the QLCA theory~\cite{Golden2010,Khrapak2018}, $\CQL$ and $\CQT$ (see Materials and Methods Eqs.~\ref{cQl} and \ref{cQt}), which are relevant for strong coupling (typically, $\varepsilon >10.7$). 
Thirdly, we compute sound velocities from the formula defining $\CQL$ and $\CQT$, except that we replaced $\wD$ by $\wA$, to obtain $\CAL$ and $\CAT$.
$\CAL$ and $\CAT$ provide a good estimate of the sound velocity related to the structure of the 2D granular assembly at scales of order the particle separation.
We find that the acoustic propagation of longitudinal waves, measured by $\CGL$, occurs at roughly 25~mm\,s$^{-1}$, that is, nearly three times the thermal velocity $3\,v_\textrm{th}=3\,\sqrt{\Ek/m}$ and the predicted adiabatic sound velocity for hard disks with dissipative collisions, which is $2.04\, v_\textrm{th}$~\cite{Brito2013} (figure ~\ref{cplot}c).
We note that both $\CGL$ and $\CGT$ remain much smaller than $\CAL$ and $\CAT$ when $\varepsilon > 10.7$ due to the cut-off frequencies.
We also observe that for $\varepsilon \gtrsim 100$ (resp., $\varepsilon >10.7$), $\CAL$ ($\CAT$) is proportional to the applied magnetic field $B$, in agreement with the scaling of $\wD$ (figure ~\ref{cplot}c,d insets) and are of the order of the theoretical values $\CQL$ and $\CQT$ for a 2D array of dipoles.

The widths of the experimental dispersion relations, $\Lambda_\textrm{L} (k)$ and $\Lambda_\textrm{T} (k)$, quantify the excitation lifetimes due to dissipation and nonlinear effects in the interaction potential, and can be considered as dimensionless decay rates when rescaled by an angular frequency. {The dissipation is caused by inelastic collisions between particles and between the particle and the top and bottom walls. Even when the particle collisions are suppressed, the level of dissipation remains significant, as shown in one of our previous work~\cite{Castillo2020}}. We show these decay rates, obtained by fitting the spectra (Eq.~\ref{doubleLorentz}) and averaged within $0<k\,a<1.68$ for each $\varepsilon$, as functions of $\varepsilon$ for both the longitudinal and transverse waves in figure ~\ref{cplot}E. We find that the decay rates equal at most roughly~1, and are closer to~0.1 in the solid crystal phase. Therefore, wave propagation is strongly damped because the excitations disappear on a time scale similar to their period.

To make a similar analysis in space instead of time, we define coherence lengths of the excitations, $\mathcal{L}_\textrm{L}$ and $\mathcal{L}_\textrm{T}$, as the ratios $\CGL/\Lambda_\textrm{L}$ and $\CGT/\Lambda_\textrm{T}$, respectively. We find the coherence lengths to be roughly between 1--2~mm, \textit{i.e.}, very close to the particle diameter, $a=1$~mm (figure ~\ref{cplot}f). Thus, the waves propagate in a strong damping regime and are not affected by the lateral boundaries of the cell, whose sides are 90~mm-long. Such strong excitation damping likely is the reason why it is difficult to detect the waves self-generated by mechanical agitation in granular media. Indeed, strongly damped density waves have been reported in only one experimental and numerical study~\cite{Clerc2008} in the context of the transient formation of solid-like clusters in a dense granular layer without interactions {and vertically shaken}.

\section{Elastic coefficients}

Now, we  non-intrusively estimate the elastic coefficients---the bulk and shear moduli---of our quasi-2D, isotropic granular medium using two methods.
% first method
In the first method, we use an extension of the QLCA theory to compute the so-called instantaneous bulk and shear moduli, $K_{\infty}$ and $G_{\infty}$, from the structural arrangement of particles and the interaction potential~\cite{Zwanzig1965,Khrapak2018,Khrapak2018c}, according to Eqs.~\ref{Kinf} and \ref{Ginf}. 
% second method
In the second method, we estimate the elastic coefficients using the sound velocities deduced from the experimental dispersion relations. Specifically, we use results from the hydrodynamics theory of granular gases---which introduces effective elastic and dissipation coefficients at scales larger than few particle separations---and from linear elasticity. In the fluid phase, the hydrodynamics theory of granular gases predicts longitudinal waves only and relates their velocity, $C_\textrm{L}$, to the bulk modulus, $K$, through $C_\textrm{L}=\sqrt{K/\rho}$, with $\rho=m N/S$ the 2D mass density, where $N$ is the number of particles inside the region of interest. In the solid phase, linear elasticity predicts $C_\textrm{L}=\sqrt{(K+G)/\rho}$ and $C_\textrm{T}=\sqrt{G/\rho}$ for the transverse waves. Therefore, consistently with both theories, we take $K=\rho\,(C_\textrm{L}^2-C_\textrm{T}^2)$ and $G=\rho\,C_\textrm{T}^2$ throughout all four regimes of our granular system. We obtain the elastic coefficients from the QLCA theory, $K_\textrm{Q}$ and $G_\textrm{Q}$, by substituting $C_\textrm{L}$ and $C_\textrm{T}$ with $\CQL$ and $\CQT$, respectively; and we deduce experimentally adjusted elastic coefficients, $K_\textrm{A}$ and $G_\textrm{A}$, similarly from the fitted sound velocities $\CAL$ and $\CAT$.

\begin{figure}[t!]
\begin{center}
\includegraphics[width=1\linewidth]{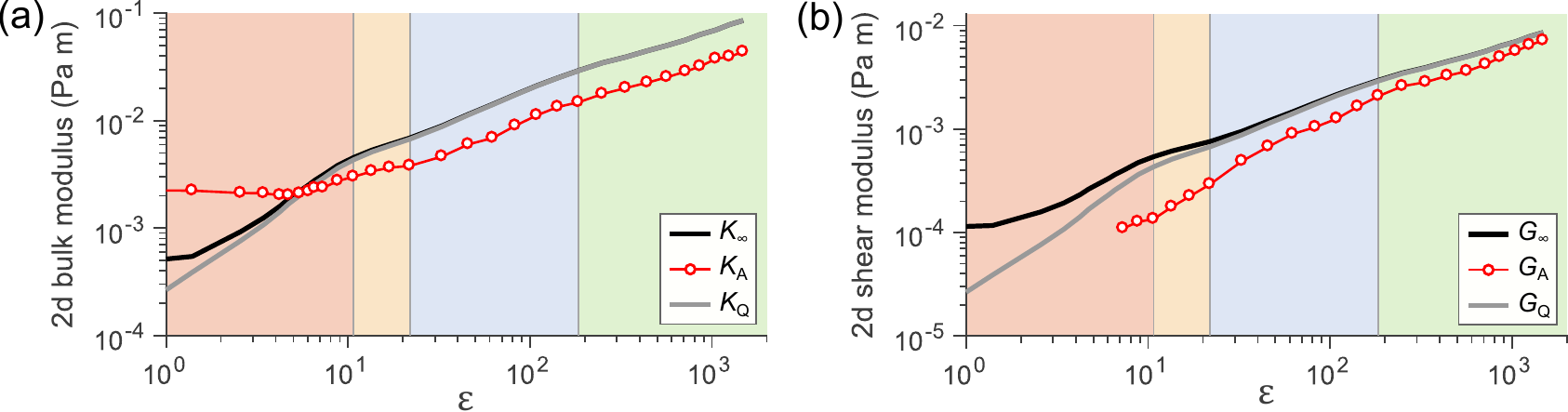}
\caption{{
2D elastic coefficients. 
(a)~Bulk modulus and (b)~shear modulus as functions of $\varepsilon$.
The solid black curves show the ``instantaneous'' bulk and shear moduli, $K_\infty$ and $G_\infty$, which are computed from the pair correlation function and the magnetic interaction potential (see Eqs.~\ref{Kinf} and \ref{Ginf}).
The red circles show $\KA$ and $\GA$, the moduli estimated within the hydrodynamics description of granular gases, from $\CAL$ and $\CAT$.
The solid grey curves show $\KQ$ and $\GQ$, the moduli obtained with $\CQL$ and $\CQT$, which are derived from the QLCA theory.
}
}
\label{Bulkshearmodulus}
\end{center}
\end{figure}

All three of our bulk modulus estimates increase with $\varepsilon$,  demonstrating the increasing rigidity of the granular assembly as magnetic interactions become stronger (figure ~\ref{Bulkshearmodulus}a).
Specifically, $K_\textrm{Q}$ and $K_{\infty}$ are nearly equal in the regimes of strong coupling ($\varepsilon > 10.7$), which reflects the consistency of the method as the velocities deduced from the QLCA theory rely on physical principles equivalent to those we use to determine  the instantaneous elastic coefficients. $K_\textrm{A}$, which is the relevant experimental bulk modulus estimate in the regimes of strong coupling, shows a similar behaviour, although with smaller values for $\varepsilon > 10.7$.

We also find that our shear modulus estimates grow with $\varepsilon$, indicating an increasing transverse stiffness in the system (figure ~\ref{Bulkshearmodulus}b).
All three bulk moduli take similar values in the solid phase. We note that, similarly to $K_{\infty}$ and $K_\textrm{Q}$, $G_{\infty}$ and $G_\textrm{Q}$ are nearly equal when the coupling is strong. For $\varepsilon < 10.7$, $G_{\infty}$ measures the visco-elastic response of the fluid phase. In contrast, $G_\textrm{A}$ vanishes in the absence of transverse waves ($\varepsilon<7$), thus showing the typical response of a fluid, in which shear excitations are damped due to viscosity and do not propagate. In the dipolar fluid regime, $G_\textrm{A}$ is significantly smaller than $G_{\infty}$ and $G_\textrm{Q}$. 

Finally, let us point out that the high-frequency moduli $K_{\infty}$ and $G_\infty$ convey the mechanical response of a static ensemble of particles to stress. In particular, $G_\infty$ can be expected to be nonzero in the granular gas regime. Indeed, in a fluid phase, $G_\infty$ measures the instantaneous viscoelastic response, whereas the shear modulus reaches zero at a time given by the viscous relaxation of shear stress.

To  summarize, the analysis of the mechanical waves that we experimentally study here enables to non-intrusively estimate the elastic coefficients of the granular system. We find elastic coefficient values within $[10^{-5}; 10^{-1}]$~Pa\,m, which is consistent with the theoretical estimates from the thermodynamics of interacting dipoles. This confirms the relevance and accuracy of the wave spectroscopy method presented here.

\section{Conclusions}
In this article, we have shown experimentally that particle velocity fluctuations in quasi-two-dimensional, driven granular media correspond to the propagation of mechanical waves in the presence of significant energy dissipation. The random motions of the particles are self-organized into collective modes, which are the macroscopic counterpart of the phonons of condensed matter. When dipolar particle interactions are weak or turned off, we only observe longitudinal, non-dispersive, compression waves---as expected for a fluid phase. This study constitutes the first step in investigating at more depth the physics of granular gases through their collective dynamics, in relation to fluctuating hydrodynamics with dissipative collisions 
~\cite{Vollmayr2011,Gradenigotheo2011,Brito2013}.
In contrast, when the coupling due to the dipolar interactions is dominant, propagating waves are both longitudinal and transverse, as well as dispersive. The experimental dispersion relations are satisfyingly modelled using the spatial distribution of particles and their interaction potential. Notably, a specificity of our experimental system is the presence of a cut-off frequencies at $k=0$. We interpret these cut-offs as optical dispersion relations arising from the coupling between horizontal motions and small vertical oscillations, the latter of which we do not have the ability to measure. For all coupling strengths, we compute characteristic sound velocities, which we use to non-intrusively estimate the elastic parameters of the granular medium.

In our system, in-plane mechanical forcing---the horizontal bead velocities acquired from rebounds on the rough bottom surface of the cell---is random, thus is non-coherent and has no characteristic frequency.
We demonstrated that, in response to this forcing, strongly damped mechanical waves propagate through the system. 
This result agrees with the linearised hydrodynamic theory of particulate systems, which predicts wave propagation as the first-order mechanical response to random forcing, for both driven granular media without remote interactions~\cite{Vollmayr2011,Brito2013} and interacting molecules~\cite{HansenMcdonaldBook}. As an extension of our system which we would suggest to implement next, a piston could be added at a cell boundary to generate either large-scale sinusoidal or pulsed mechanical waves into the granular medium~\cite{Buttinoni2017}, to bridge the gap between the large-scale sound waves and the incoherent random waves that we evidenced in this driven, interacting granular material. 
Another interesting direction would be to compare the large scale elastic coefficients with those obtained here, at the particle level, by subjecting the granular quasi-monolayer to an external mechanical perturbation, such as shear or compression.\\

\section{Methods}
\subsection*{Experimental setup, protocol, and image analysis} 
The granular particles are chromed alloy steel (AISI 52100) spheres of diameter $a=1$\,mm and of mass $m=4.07$\,mg. They are ferromagnetic with high magnetic permeability and a low remnant magnetic field, so that their coercive field is small compared with the values of the magnetic field $B$ used in our experiments. A total of $N_0 = 2000$ particles are placed in a square aluminium cell with sandpaper covering its bottom surface, and sealed by a smooth, transparent acrylic lid. The area between the aluminium walls is $S_0 = 9$~cm $\times$ 9~cm and the gap between the rough bottom surface and the lid is $1.42\,a$. {To avoid any electrostatic effects, the top transparent lid is made of polycarbonate with an antistatic coating. Linked to an electric ground, this lid releases the electrostatic charges that may be temporarily carried by particles.} The cell is mounted horizontally on and vibrated vertically by an electromagnetic shaker. The mechanical forcing is sinusoidal, of frequency 300~Hz, and with a root mean square acceleration of $\Gamma=1.6\,g$, with $g$ the  gravitational acceleration. A pair of coils generate a nearly homogeneous vertical magnetic field across the cell area (the magnetic field is roughly 3\% smaller at the edges of the cell than at its center). Immersed in this magnetic field, the particles are magnetized into induced, vertical dipoles. The average area fraction of particles is $\phi_0=\frac{N_0\,\pi a^2}{4S_0} \approx 0.2$.

An annular light-emitting diode (LED) array is centred above the cell and produces a high-contrast circular signal by reflection on the chromed particles. We record pictures of the particle assembly from above using a high-speed video camera (Phantom V10) at high frequency (780 Hz) and resolution ($1152\times1152$\,pixels). The window of observation around the cell center is of area $S=50.36$ $\times$ 50.36~mm$^2$. One particle diameter corresponds to roughly $20$\,pixels.

We determine the positions of particle centres using  a convolution-based least-square fitting routine~\cite{Reis2007} completed by an intensity-weighted center detection algorithm (accuracy estimated to better than $0.3$\,pixel). The trajectories and velocities of each particle projected in  the horizontal plane are reconstructed using a tracking algorithm~\cite{Crocker1996} in the window of observation ${S}$. 

In each experimental run, first, we raise the amplitude of the mechanical shaker to $1.6g$ and the magnetic field to the desired value, second, we wait an equilibration time of $100$\,s, and finally, we record 3000 frames using the high-speed camera during $3.85\,$s. For each value of $\varepsilon$, we run five independent and identical experiments; the data that we present in this article is obtained from their average.

The area fraction of particles within $S$ is computed for each run as $\phi=\frac{N\,\pi a^2}{4S}$. Note that due to inelastic clustering at small $\varepsilon$ and non-repelling boundaries, $\phi$ decreases from $0.34$ to $0.17$ across the full range of $\varepsilon$~\cite{Merminod2014,Castillo2020}.

\subsection*{Magnetic potential energy, kinetic energy, and hexagonal order parameter}

From the positions and velocities of the $N$ particles that we track, we compute the mean magnetic energy per particle as
\begin{equation}
    \Em=\left\langle \,\frac{1}{N}\sum_{i=1}^N \sum_{j=i+1}^N U_{i,j} {(r_{i,j})} \right\rangle\,, 
    \label{Em}
\end{equation}
and the mean kinetic energy per particle as
\begin{equation}
    \Ek=\left\langle \frac{m}{2N}\sum_{i=1}^N v_i^2 \right\rangle
    \label{Ek}
\end{equation}
with $r_{i,j}$ the distance between particles $i$ and $j$ in the horizontal plane, $U_{i,j}$ their interaction potential (Eq.~\ref{Uij}), $v_i$ the velocity of particle $i$, and the brackets denoting time averaging.

We monitor the degree of hexagonal order through the sixfold bond-orientational order parameter $\Psi_6^j$, also called the local hexagonal order parameter,
\begin{equation}
\Psi^j_6=\frac{1}{n_j}\sum_{k=1}^{n_j}\mathrm{e}^{6 i \theta_{jk}},
\end{equation}
where $n_j$ is the number of nearest neighbours of particle $j$, and $\theta_{jk}$ is the angle between the neighbour $k$ of particle $j$ and a reference axis. By definition, $\Psi_6^j=1$ when the nearest neighbours form a regular hexagon. The corresponding global average,
\begin{equation}
|\Psi_6| = \left |\left \langle \frac{1}{N}\sum_{j=1}^N \Psi^j_6\right\rangle\right|,
\label{Psi6b}
\end{equation}
where the vertical bars denote a modulus, measures the degree of hexagonal order of the particle assembly. We refer to $|\Psi_6|$ as the global hexagonal order parameter of the system.

\subsection*{Current correlation functions}

The velocities of individual particles are analysed in the spatial Fourier space. With the hypothesis of homogeneous and isotropic system, according to the linearised hydrodynamics theory of molecular fluids~\cite{HansenMcdonaldBook}, the coarse-grained velocity field can be inferred from the computation of longitudinal current correlation function and of the transverse current correlation function. This approach has been extended to the case granular systems, for which energy injection and dissipation are also included~\cite{vanNoijeErnst1999,Vollmayr2011,Gradenigotheo2011,Brito2013}. Here, we define in 2D, the  longitudinal current correlation function $J_\mathrm{L} (\mathbf{k},t)$ and the transverse current correlation function $J_\mathrm{T} (\mathbf{k},t)$, as~\cite{HansenMcdonaldBook,CastilloPhd,Puglisi2012}
\begin{equation}
J_\textrm{L} (\mathbf{k},t)=\left\langle \frac{1}{N}  \sum_{i,j=1}^N (\mathbf{\hat{k}} \cdot \mathbf{v}_i (t)) \, (\mathbf{\hat{k}} \cdot \mathbf{v}_j (0))\, \mathrm{e}^{\imath\,\mathbf{k}\, (\mathbf{r_i (t)-r_j(0))} }    \right\rangle
\end{equation}
and
\begin{equation}
J_\textrm{T} (\mathbf{k},t)=\left\langle \frac{1}{N}  \sum_{i,j=1}^N (\mathbf{\hat{k}} \times \mathbf{v}_i (t)) \, (\mathbf{\hat{k}} \times \mathbf{v}_j (0))\, \mathrm{e}^{\imath\,\mathbf{k}\, (\mathbf{r_i (t)-r_j(0))} }    \right\rangle 
\end{equation}
where $\mathbf{\hat{k}}$ is the unitary vector directed along the wave vector $\mathbf{k}$. These functions are computed on a $150 \times 150$ with node values $0.0624\,$mm$^{-1} < k_{x,y} <9.36\,$mm$^{-1}$. {Each pair $(k_x,k_y)$ is discretized according to $(n_x\pi/L_x,n_y\pi/L_y)$, where $n_x, n_y\in\mathbb{N}$ and $L_x=L_y=50.36$\,mm.} The brackets $ \left\langle \right\rangle $ denote time averaging.

Then assuming system isotropy, we define the angle-averaged current correlation functions,
\begin{equation}
    J_\mathrm{\alpha} (k,t)=(2\pi)^{-1} \,\int_0^{2\pi} J_\mathrm{\alpha} (\mathbf{k},t)\, \mathrm{d} \theta \, , \label{Jlkt}
\end{equation}
where the subscripts $\alpha$ should be substituted by ``L'' for longitudinal waves, and ``T'' for transverse waves. These functions are also called dynamical structure factors of longitudinal and transverse velocity modes~\cite{Gradenigotheo2011,Puglisi2012}. Finally, $J_\mathrm{\alpha} (k,t)$ is averaged over five independent runs with identical experimental parameters.

\subsection*{Space-time spectra}

According to the Wiener-Khinchin theorem, the space-time spectra of excitations $S_\mathrm{\alpha}(k,\omega)$ are directly deduced from the temporal current correlation functions by a time Fourier transform as
\begin{equation}
    S_\mathrm{\alpha}(k,\omega)=\mathrm{Re}\,\, \int_0^\infty J_\mathrm{\alpha} (k,t) \,\mathrm{e}^{\imath \,\omega\,t}\,  \mathrm{d} t .
    \label{spectraSvl} 
\end{equation}
Here, the time Fourier transform is performed using the fast Fourier Transform algorithm and a Hann window as tapering function. The frequencies are discretised into 2999 values between 0~Hz and 390~Hz.

To deduce experimental dispersion relations from the space-time spectra, we assume that the current correlations of propagating waves can be modelled as damped oscillations, that is, 
\begin{equation}
    J_\mathrm{\alpha} (k,t) \propto \mathrm{e}^{-\Lambda_\mathrm{\alpha} (k)\, t}\, \cos (\omega_\mathrm{R,\alpha} (k)\,t)\, ,
    \label{dampedcos}
\end{equation}
which, using Eq.~\ref{spectraSvl}, leads to the space-time spectra being modeled as proportional to double Lorentzian functions~\cite{Khrapak2018,Mistryukova2019},
\begin{eqnarray}
 S_\mathrm{\alpha} (k,\omega) \! \propto \!  \dfrac{1}{(\omega -\omega_\mathrm{R,\alpha} (k))^2 
 +\Lambda_\mathrm{\alpha}^2 (k)} \!
  + \! \dfrac{1}{(\omega +\omega_\mathrm{R,\alpha} (k))^2+\Lambda_\mathrm{\alpha}^2 (k)}
\label{doubleLorentz}
\end{eqnarray}
where $\omega_\mathrm{R,\alpha} (k)$ is the fitted experimental dispersion relation, and $\Lambda_\mathrm{\alpha}(k)$ is the fitted width of the dispersion relation, which quantifies the wave damping. In practice, for each value of wavenumber $k$, we fit our experimental time-space spectra with Eq.~\ref{doubleLorentz} to obtain $\omega_\mathrm{R,\alpha} (k)$ and $\Lambda_\mathrm{\alpha}(k)$.

\subsection*{QLCA theory for a 2D array of interacting dipoles}
\label{QCLA}
To describe the collective excitations in a 2D system of interacting dipoles in strong coupling regime for fluid and solid phases, Golden \textit{et al.}~\cite{Golden2008,Golden2010} proposed the quasilocalized charge approximation (QLCA) method. It extends the usual calculation of the phonon dispersion relation in a lattice within the harmonic approximation, by taking into account the actual average position of particles through the input of the pair correlation function $g(r)$ in the continuous limit and assuming isotropy. Khrapak \textit{et al.}~\cite{Khrapak2018} provided explicit expressions of the dispersion relations, using the Wigner-Seitz radius $a_\textrm{WS}$ as a characteristic distance:
  \begin{eqnarray}
\omega_\textrm{Q,L}^2&=& \dfrac{3\wD^2}{2}\,\int_0^\infty\, \dfrac{g(x)\,\mathrm{d} x}{x^4}\,[3-J_0(q\,x)+5\,J_2(q\,x)] \label{eqwQL}\\
\omega_\textrm{Q,T}^2&=& \dfrac{3\wD^2}{2}\,\int_0^\infty\, \dfrac{g(x)\,\mathrm{d} x}{x^4}\,[3-J_0(q\,x)-5\,J_2(q\,x)]
\label{eqwQT}
\end{eqnarray}
with $J_0(x)$ and $J_2(x)$ the Bessel functions of the first kind, $x=r/a_\textrm{WS}$ the reduced distance, $q=k\,a_\textrm{WS}$ the reduced wavenumber, and $\wD$ the characteristic dipole oscillation angular frequency. The Wigner-Seitz radius $a_\textrm{WS}$ is directly related to the particle surface density $\rho_{\mathrm{N}}=N/S$, by $a_\textrm{WS}=(\pi\,\rho_{\mathrm{N}})^{-1/2}$. From the QCLA theory, $\wD$ reads:
 \begin{equation}
\wD=\left(\dfrac{2\pi\,\rho_{\mathrm{N}}}{m\,(a_{\mathrm{WS}})^3} \, \dfrac{4\,\pi\,B^2}{\mu_0}\,\left(\dfrac{a}{2}\right)^6 \right)^{1/2}\, .
\label{omegaD}
\end{equation}
Knowing that in our system $a_{\mathrm{WS}}=a/(2\sqrt{\phi})$ and $\rho_{\mathrm{N}}=4\phi/(\pi\,a^2)$, the characteristic frequency can be expressed as a function of the experimental parameters:
 $$\wD=B\,\phi^{5/4}\,\sqrt{\dfrac{4\pi\,a}{\mu_0\,m}}\,.$$
In the limit $k\rightarrow 0$, the QLCA theory predicts an acoustic behaviour, \textit{i.e.}, a non dispersive wave propagation, $\omega  \propto k$, with the longitudinal and transverse sound velocities respectively:
\begin{eqnarray}
C_\textrm{Q,L}=\sqrt{\dfrac{33\,M}{16}}\,\wD\,a_{\mathrm{WS}}\, \int_0^\infty \dfrac{g(r)}{r^2} \mathrm{d} r \label{cQl} \\
C_\textrm{Q,T}=\sqrt{\dfrac{3\,M}{16}}\,\wD\,a_{\mathrm{WS}} \, \int_0^\infty \dfrac{g(r)}{r^2} \mathrm{d} r  \,,
\label{cQt}
\end{eqnarray}
with $M=(1/2)\,\sum_i (r_i/a_{\mathrm{WS}})^{-3}\approx 0.798512$ the Madelung constant for a perfect hexagonal lattice. For the set of experiments presented here, the experimental distribution of the particles in the different regimes gives: $ \int_0^\infty {g(r)}\,r^{-2}\, \mathrm{d} r \approx 0.85$. 

Finally, the bulk and shear moduli obtained from the QLCA theory, $K_{\infty}$ and $G_{\infty}$, are given by:
\begin{eqnarray}
K_{\infty} \! = \! 2\,\rho_{\mathrm{N}}\,\Ek- \dfrac{\pi\,\rho_{\mathrm{N}}^2}{4} \int_0^{\infty}\,r^2\,\left[\dfrac{\mathrm{d} U (r)}{\mathrm{d} r}-r \,\dfrac{ \mathrm{d}^2 U (r)}{\mathrm{d} r^2}\right]\,g(r)\,\mathrm{d} r \label{Kinf} \\
G_{\infty} \! = \! \,\rho_{\mathrm{N}}\,\Ek + \dfrac{\pi\,\rho_{\mathrm{N}}^2}{8} \int_0^{\infty}\,r^2\,\left[\dfrac{3\,\mathrm{d} U (r)}{\mathrm{d} r}+r \, \dfrac{\mathrm{d}^2 U (r)}{\mathrm{d} r^2}\right]\,g(r)\,\mathrm{d} r  \label{Ginf}
\end{eqnarray}
These moduli are qualified of high frequency, or instantaneous, because they characterize the mechanical response for a static arrangement of particles in space.

\vskip6pt

\enlargethispage{20pt}

%\ethics{Insert ethics text here.}
\enlargethispage{20pt}

%\ethics{Insert ethics text here.}

\textit{{The data corresponding to the article 's figures are available at: \url{http://www.msc.univ-paris-diderot.fr/~berhanu/DataRepositoryBerhanuRSPA2022.html}}} \\

\textit{{M.B.: conceptualization, formal analysis, funding acquisition, software, resources, writing original draft; S.M.: conceptualization, investigation, formal analysis, software, writing review and editing; G.C.: conceptualization, funding acquisition, formal analysis, software, writing review and editing; E.F.: conceptualization, funding acquisition, resources, writing review and editing.} }\\

\textit{{This research was supported by {Universit\'e Paris Cit\'e} and Fondecyt Grant No. 11200464 (G.C.).}}\\

\begin{acknowledgments}{We thank at MSC, {Universit\'e Paris Cit\'e}, Thierry Hocquet and Martin Devaud for granting us access to their computing facilities; and Marc Durand, Michel Saint-Jean, Christophe Coste and François Gallet for discussions. We acknowledge Nicol\'as Mujica and Rodrigo Soto from University of Chile for discussions.}\\
\end{acknowledgments}

%\disclaimer{Insert disclaimer text here.}

%%%%%%%%%% Insert bibliography here %%%%%%%%%%%%%%

\vskip2pc

%\noindent {\bf Please follow the coding for references as shown below.}
%
%%\begin{thebibliography}{9}
%%
%%\bibitem{1} Allwood JM, Cullen JM. 2011 \textit{Sustainable materials:  with both eyes open}.
%%Cambridge, UK: UIT Cambridge. See \href{http://www.withbotheyesopen.com}{http://www.withbotheyesopen.com}.
%%
%%\bibitem{2}  MacKay DJC. 2008  \textit{Sustainable energy:  without the hot air}.
%% Cambridge, UK: UIT Cambridge. See \href{http://www.withouthotair.com}{http://www.withouthotair.com}.
%%
%%\bibitem{3} Gallman PG. 2011  \textit{Green alternatives and national energy strategy: the facts
%% behind the headlines}.  Baltimore,\ MD: Johns Hopkins University Press.
%%
%%\bibitem{4} MacKay DJC. 2013.  Solar energy in the context of energy use, energy transportation, and
%% energy storage. \textit{Proc. R. Soc. A} \textbf{371}.
%%
%%\end{thebibliography}
%
%\noindent If maintaining .bib file for references, then please use "RS.bst" to generate the references.
%
%\noindent Example:

%\bibliographystyle{jfm}

%\bibliography{Biblio_Berhanuetal2021} %%%%% .Bib file



\section*{Supplementary material}

# Supplementary Information for
# Wave spectroscopy in a driven granular material

Michael Berhanu[1], Simon Merminod[2], Gustavo Castillo[3] and Eric Falcon[1]

[1] Université Paris Cité, CNRS, MSC, UMR 7057, F-75013, Paris, France

[2] Department of Molecular and Cellular Biology, Harvard University, Cambridge, MA 02138, USA

[3] Instituto de Ciencias de la Ingeniería, Universidad O'Higgins, 2841959 Rancagua, Chile

## Supporting Information Text

In this appendix, we provide first in Section 1, typical top view images of the four regimes, extracted from the corresponding movies in Supplemental material. These images characterize the four regimes defined in the main text. In Section 2, we show a schematic and a picture of the experimental device. In Section 3, we discuss the influence of three-dimensional (3D) effects in the propagation of mechanical waves in our experimental system. In Section 4, we propose a simple phenomenological model to explain qualitatively the coupling between a horizontal wave with a vertical oscillation, in order to explain the observation of cut-off frequencies in the experimental dispersion relations. In Section 5, the pair correlation function $g(r)$ is plotted in order to better characterize the four regimes studied in this work. Moreover, this function $g(r)$ is used as an input to apply the QCLA theory to our measurements. Finally, in Section 6, we show the mean-square displacements as a function of time for the four regimes previously defined in the main text.

# 1. Illustration images and movies of the different regimes

Movies S1 to S4. For each regime, top view of the system in the windows of the observation $S = 50.36 \times 50.36$ mm$^2$. The scrolling speed is slowed down by a factor of 5.

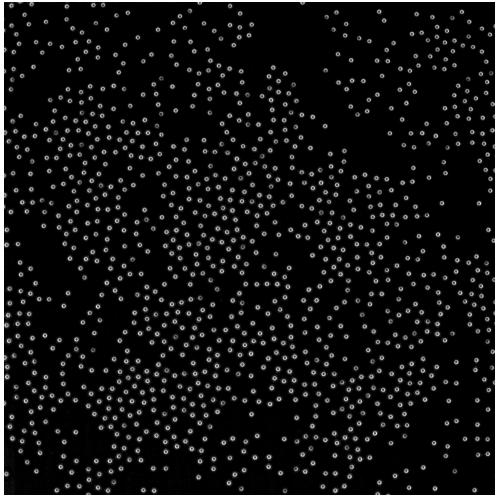

FIGURE 1. Movie S1. Granular gas. $\varepsilon = 0$, $B = 0$ G

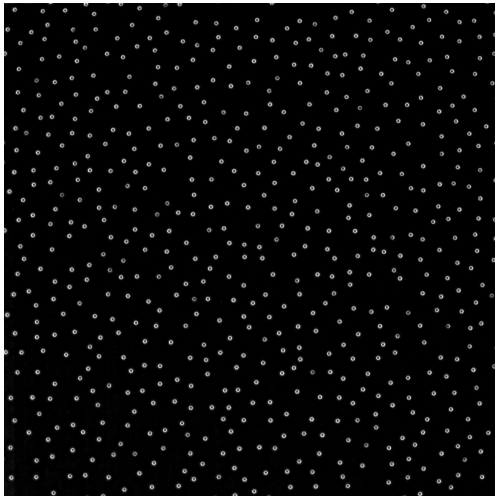

FIGURE 2. Movie S2. Dipolar fluid. $\varepsilon = 17$, $B = 93$ G.

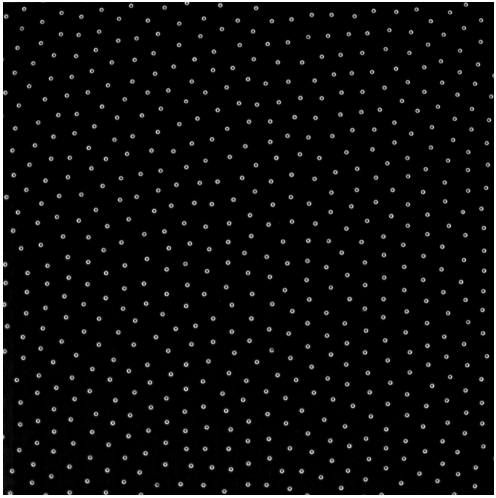

FIGURE 3. Movie S3. Partial order. $\varepsilon = 83$, $B = 185$ G.

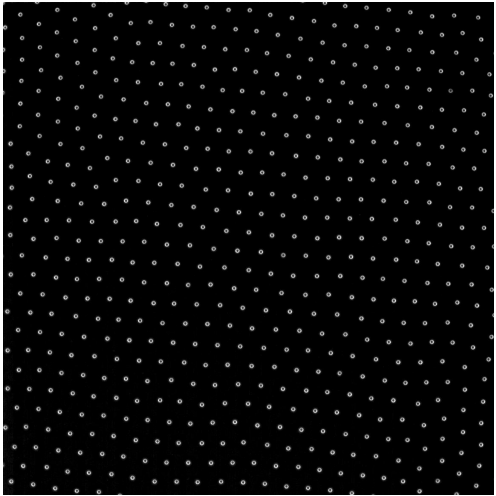

FIGURE 4. Movie S4. Solid crystal. $\varepsilon = 717$, $B = 349$ G.

Movies S5 to S8. In order to better visualize the motions, we also provide movies in slow motion and with higher magnification for each regime. The window of observation is now $26.23 \times 26.23$ mm$^2$. The scrolling speed is slowed down by a factor of 30.

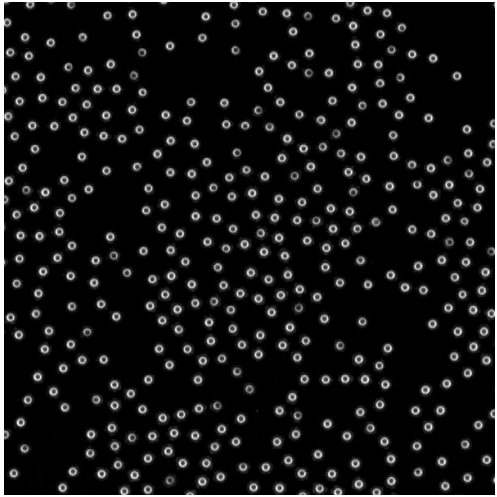

FIGURE 5. Movie S5. Granular gas. $\varepsilon = 0$, $B = 0$ G

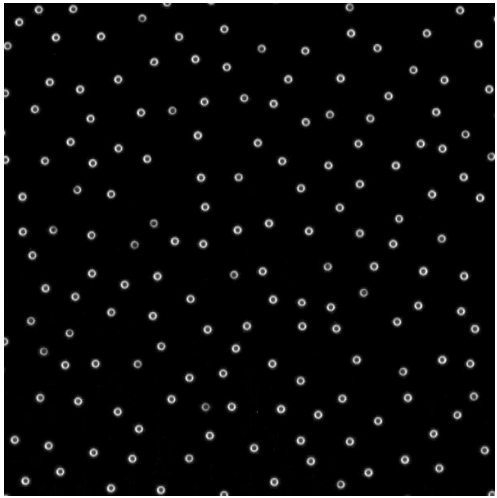

FIGURE 6. Movie S6. Dipolar fluid. $\varepsilon = 17$, $B = 93$ G.

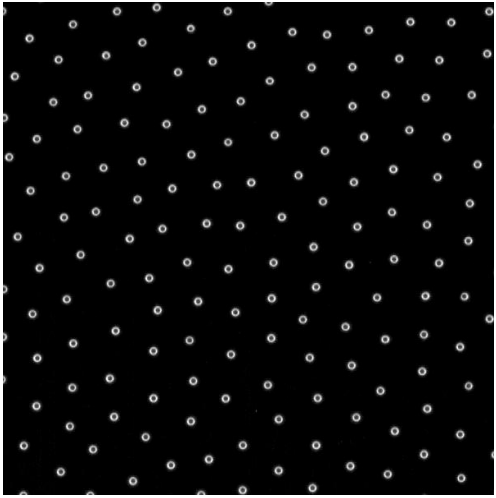

FIGURE 7. Movie S7. Partial order. $\varepsilon = 83$, $B = 185$ G.

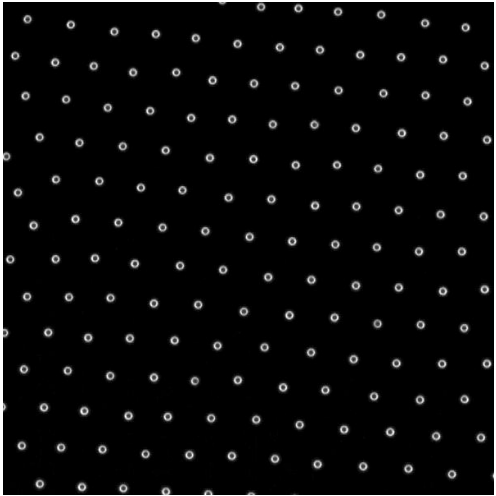

FIGURE 8. Movie S8. Solid crystal. $\varepsilon = 717$, $B = 349$ G.

## 2. Schematic of the experimental device

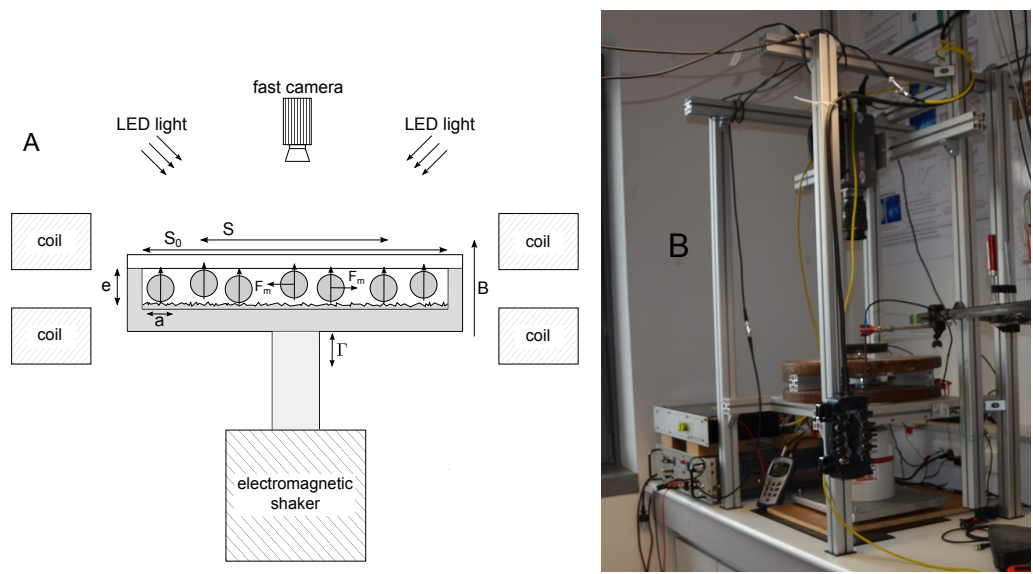

FIGURE 9. (A) Schematic of the experimental setup (see *main text Section 6 Methods* for description). (B) Picture of the experimental device.

## 3. Discussion of 3D effects and order of magnitude of the vertical oscillation frequency.

In the main text of the article, the analysis is essentially performed in 2D, in the horizontal plane, in which the positions of the particles are measured. This assumption is justified by the vertical confinement of the beads between two horizontal plates with a gap equal to $e = 1.42\,a$ with $a$ the particle diameter. Two particles separated by a distance $r$ in the cell are not exactly in the same plane and deviation to the two dimensionality may be taken into account, specifically for the dipole-dipole interaction potential between the particles. Magnetized soft ferromagnetic spheres behave indeed as induced dipoles, whose magnetic moments are vertical and proportional to $B$. In the limit of high magnetic permeability, the interaction potential $U_m$ of a particle located at a distance $r$ and a polar angle $\theta$ from a second particle, reads in spherical coordinates, with $\mu_0$ the vacuum permeability Jackson (1999); Merminod *et al.* (2015):

$$U_m(r, \theta) = -\frac{\pi}{16} \frac{B^2}{\mu_0} \frac{a^6}{r^3} \left(3\cos^2\theta - 1\right) . \tag{3.1}$$

This potential has a maximum in $\theta = \pi/2$. Two spheres in the same horizontal plane are thus repelling each-other. The interaction remains repulsive for $\arccos(1/\sqrt{3}) < \theta < \pi - \arccos(1/\sqrt{3})$. For the set of experiments presented here with $e = 1.42\,a$, $\theta$ remains close to $\pi/2$. In the most unfavorable case, when two spheres are in contact, one touching the bottom and the other the lid, $\theta$ is constrained in the interval $[\arccos(e/a - 1), \pi - \arccos(e/a - 1)] \approx [1.14, 2.00]$. The particles are thus always repelling each-other. However, for two particles separated by a given distance $r$ the state of minimal energy corresponds to the maximal accessible angular deviation to $\theta = \pi/2$, *i.e.* $\theta_{m1} = \arccos((e - a)/r)$ or $\theta_{m2} = \pi - \arccos((e - a)/r)$. Therefore, for negligible agitation, the particles on a line should form an alternation of up and down particles, slightly above and below the mid-plane as illustrated in Fig. 10. For such a zigzag structure in a plane but with a repulsive electrostatic potential, it has been shown that the propagation of mechanical waves involve dispersion relations with acoustic and optical branches Dessup *et al.* (2015). However, the extension in 3 dimensions is more complex even in the crystal regime. The up and down alternation is indeed not compatible with a triangular lattice, leading to a phenomenon of geometric frustration Han *et al.* (2008).

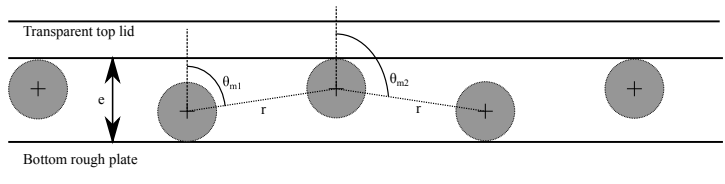

FIGURE 10. Schematic of the state minimizing the magnetic interaction potential (Eq. 3.1) for spheres distributed in the vertical plane. They form an alternation of up and down particles. However, the extension of this schematic to 3D is questionable, because the hexagonal lattice is not compatible with the up and down alternation, leading to frustrated bonds.

For this geometrical configuration, justified when the agitation is negligible in front of the magnetic interaction, *i.e.* for large $\varepsilon$, we can estimate approximately the typical frequency for vertical oscillations of particles. Considering the hexagonal lattice with a particle separation in the mid-plane $d_{hexa}$ (Eq. 5.1), the characteristic frequency for the small vertical oscillations of one particle located at the bottom of the cell can be computed numerically as $\sqrt{1/m \left(\partial^2 E_m/\partial z^2\right)}$, where $E_m$ is computed using Eq. 3.1 for the 6 nearest

neighbors forming an hexagon in the plane. We find an angular frequency $\omega_{v,1}$ close to $\omega_D$, the characteristic dipole frequency (Eq. 2 in the main text). If we consider, that the 6 nearest neighbors forming the hexagon are in the mid-plane instead of the bottom of the cell, we find a slightly larger angular frequency of oscillation $\omega_{v,2}$, even closer to $\omega_D$. The ratios $\omega_{v,1}/\omega_D$ and $\omega_{v,2}/\omega_D$ are plotted as a function of $\varepsilon$ in Fig. 11 A. These vertical oscillations frequencies are compared with the cut-off frequencies $\omega_{0,L}$ and $\omega_{0,T}$ in Fig. 11 B. We find that these cut-off frequencies have thus the same order of magnitude than the estimated vertical oscillation frequencies. We note that at high value of $\varepsilon$, $\omega_{v,1}$ becomes very close to $\omega_{0,L}$ and $\omega_{0,T}$. The origin of these cut-off frequencies may be related to the coupling of the waves measured in the horizontal plane with the small oscillations of the particles in the vertical plane.

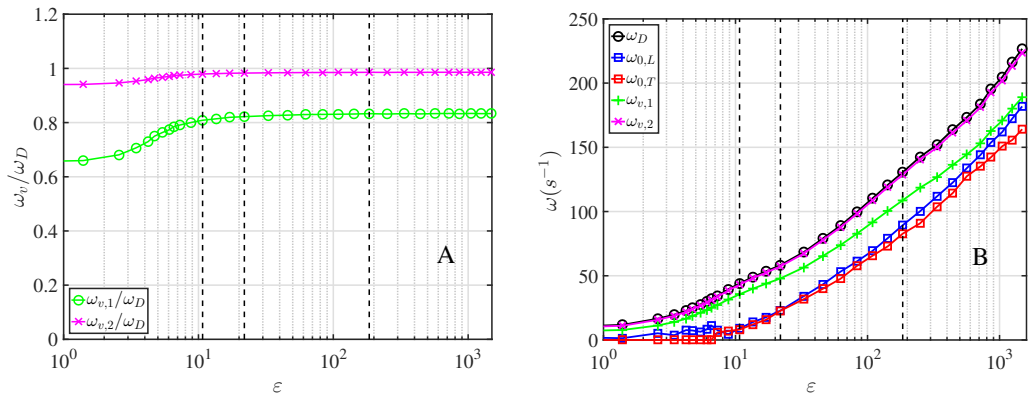

FIGURE 11. Estimations of the angular frequencies of vertical oscillations $\omega_{v,1}$ (6 neighbors located in the bottom plane) and $\omega_{v,2}$ (6 neighbors located in the mid-plane). (A) $\omega_{v,1}/\omega_D$ and $\omega_{v,2}/\omega_D$ as a function of $\varepsilon$. (B) $\omega_{v,1}$ and $\omega_{v,2}$ are compared as a function of $\varepsilon$ with $\omega_D$ and the experimental cut-off angular frequencies $\omega_{0,L}$ and $\omega_{0,T}$.

# 4. Simple model of wave propagation in a 1D mechanical system of coupled masses.

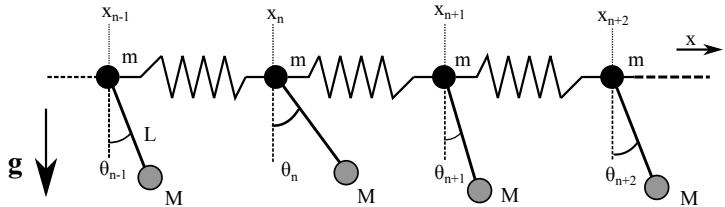

FIGURE 12. Schematic of the one-dimensional system of coupled masses. The equilibrium distance between the masses $m$ is $a$.

The coupling between a horizontal mechanical wave with a localized excitation can be simply illustrated with a phenomenological system. We consider a linear infinite chain of masses $m$ located by the coordinates $x_n$ (see schematic in Fig. 12). The mass are coupled to each-other by springs of stiffness $K$ and of length at rest $a$. Each mass is connected to a pendulum of mass $M$ and length $L$. The deviation angle to the vertical axis aligned with the gravity field $\mathbf{g}$ is noted for each pendulum $\theta_n$. Due to the masses $M$ and $m$ the variables $x_n$ and $\theta_n$ are coupled by the inertia. For small oscillations, the application of the second law of Newton gives:

$$(1 + \alpha)\,\ddot{x}_n + \omega_0^2\,(2\,x_n - x_{n+1} - x_{n-1}) = \alpha\,L\,\ddot{\theta}_n \tag{4.1}$$

$$\ddot{\theta}_n + \Omega^2\,\theta_n = -\frac{\ddot{x}_n}{L} \tag{4.2}$$

with $\alpha = M/m$, $\omega_0 = \sqrt{K/m}$ and $\Omega = \sqrt{g/L}$.
We perform then the decomposition into normal modes with

$$x_n(t) = \frac{1}{\sqrt{N}}\,\sum_k\,u_k(t)\,\mathrm{e}^{\imath\,k\,na}$$

$$\text{and}\quad \theta_n(t) = \frac{1}{\sqrt{N}}\,\sum_k\,\vartheta_k(t)\,\mathrm{e}^{\imath\,k\,na}\,.$$

Assuming the propagation of a wave whose angular frequency is $\omega$, we search solutions under the form: $u_k(t) = U_k\,\mathrm{e}^{\imath\omega t}$ and $\vartheta_k(t) = \Theta_k\,\mathrm{e}^{\imath\omega t}$. One obtains the following linear differential system:

$$\left[-(1 + \alpha)\,\omega^2 + \omega_0^2\,4\,\sin^2\left(\frac{k\,a}{2}\right)\right]\,U_k + \alpha\,L\,\omega^2\,\Theta_k = 0 \tag{4.3}$$

$$\frac{\omega^2}{L}\,U_k + (\Omega^2 - \omega^2)\,\Theta_k = 0 \tag{4.4}$$

The dispersion relation of the waves is given by the cancellation of the determinant of the linear system:

$$\left[-(1 + \alpha)\,\omega^2 + \omega_0^2\,4\,\sin^2\left(\frac{k\,a}{2}\right)\right]\,(\Omega^2 - \omega^2) - \alpha\,\omega^4 = 0 \tag{4.5}$$

The left part reminds the classic dispersion relation of a chain of mass coupled by springs, $\omega_{R,p} = 2\,\omega_0\,\sin\left((k\,a)/2\right)$. The coupling through the parameter $\alpha$ modifies the relation dispersion and allows several solutions to Eq. 4.5 for a given $k$, *i.e.* several branches of the dispersion relations.Two examples of numerical resolutions for weak

coupling are displayed in Fig. 13, A with the parameters $a = 1$, $\alpha = 0.1$, $\omega_0 = 1$ and $\Omega = 0.2$ and B with the parameters $a = 1$, $\alpha = 0.1$, $\omega_0 = 1$ and $\Omega = 0.8$. For this choice of parameters, we find two distinct branches of the dispersion relation: an acoustic branch $\omega_{R,1}$ which crosses the origin and an optical branch with a cut-off frequency which is very close to $\Omega$. Moreover, we find the relation dispersion without pendulums, $\omega_{R,p}$ fits well the acoustic branch at low $k$, then well the optical branch at large $k$. We observe also, that the optical branch can be satisfyingly approximated by an expression inspired by the Eq. 3 of the main text, $\omega_{R,c}^2 = \omega_0^2 + \Omega^2$. To conclude, this simple model shows that the coupling of a longitudinal mechanical wave with a localized oscillation modifies the dispersion relation of the longitudinal wave by creating two distinct branches, an acoustic and an optical. The optical branch with a cut-off frequency in $k = 0$ resembles the experimental dispersion relations reported in the main text (*Experimental spectra of velocity fluctuations*). This simple model provides thus a qualitative explanation of some of our observations, our experimental system being more complex. In particular, we do not observe in the experiments the signature of the longitudinal branch.

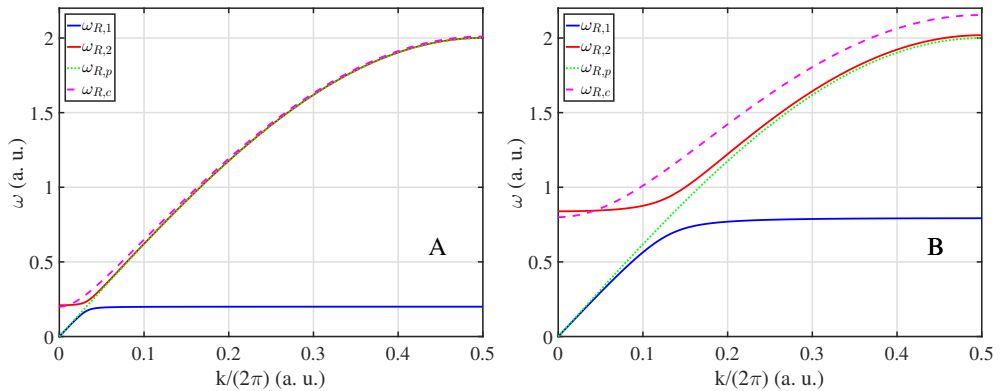

FIGURE 13. Dispersion relations for the simple mechanical model in arbitrary units. A Resolution of Eq. 4.5 for $a = 1$, $\alpha = 0.1$, $\omega_0 = 1$ and $\Omega = 0.2$. B Resolution of Eq. 4.5 for $a = 1$, $\alpha = 0.1$, $\omega_0 = 1$ and $\Omega = 0.8$. Blue $\omega_{R,1}$ first solution, acoustic branch. Red $\omega_{R,2}$ second solution, optical branch. For comparison the classic solution $\omega_{R,p}$ (green dotted line) for a chain of mass connected by springs (without the pendulums) is also provided. Finally, $\omega_{R,c}$ (magenta dashed line) is an approximate relation defined by $\omega_{R,c}^2 = \omega_0^2 + \Omega^2$.

## 5. Pair correlation function

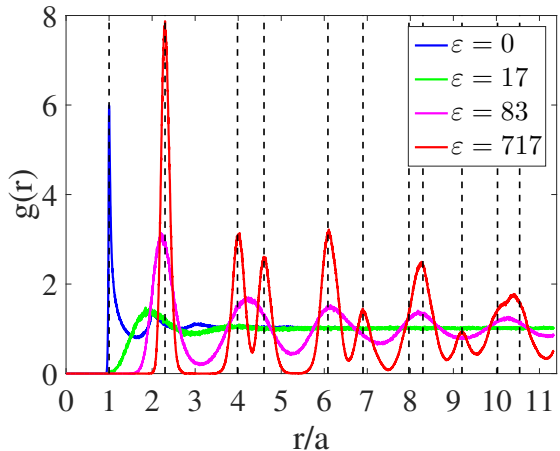

FIGURE 14. Pair correlation function $g(r)$ as a function of $r/a$ with $a$ the particle diameter for selected values of $\varepsilon$. The vertical dashed lines correspond to $r = a$ and to the expected positions of particles for a perfect hexagonal lattice (see text) at large $\varepsilon$.

The pair correlation function (radial distribution function) $g(r)$ constitutes a classic tool to visualize the changes of the structure of a particle set. This function is computed using a specific data set acquired with a frame rate of $10\,\mathrm{Hz}$ and four examples corresponding to the four regimes defined in the main text are displayed in Fig. 14. At small $\varepsilon$, one finds the characteristic shape for a dissipative granular gas, with an excluded space for $r < a$, an enhanced peak at contact for $r/a = 1$ and weak correlations for large separations, $g(r) \approx 1$ for $r \gtrsim 4a$. For larger but moderate $\varepsilon$, the peak at contact decreases as the collision rate is reduced, as described previously. An exclusion zone larger than $a$ appears, creating a first peak for $r > a$. Then, for $\varepsilon \gtrsim 20$, a succession of peaks become visible, due to the appearance of long range order. The peaks are broadened by the agitation and they become more defined for increasing values of $\varepsilon$. For $\varepsilon \gtrsim 400$, the position of the peaks correspond to those expected for a hexagonal (triangular) lattice. For the case $\varepsilon = 717$, the first peak is located at $r_{1,p} = 2.30\,a$, which is close to the lattice distance expected for a perfect hexagonal lattice

$$d_{hexa} = \sqrt{\pi/(2\sqrt{3}\,\phi)}\,a = 2.34\,a \tag{5.1}$$

Then, we find the secondary peaks as the expected multiples of the lattice distance $r_{1,p}$ for factors $\sqrt{3}$, 2, $\sqrt{7}$, 3, $2\sqrt{3}$, $\sqrt{13}$, 4, $\sqrt{19}$, $\sqrt{21}$ ... , except for the one at $2\sqrt{3}$, probably masked by the important peak at $\sqrt{13}$.

## 6. Mean square displacements

Another method to characterize structural and dynamical changes consists in measuring the mean square displacements (MSD) of the particles, defined by $\mathrm{MSD}(t) = \langle |\mathbf{r}_i(t) - \mathbf{r}_i(0)|^2 \rangle_i$, where $\mathbf{r}_i(t)$ is the position of the particle $i$ at the time $t$, $t_0$ being an arbitrary time origin. The MSD are displayed for selected values of $\varepsilon$ in Fig. 15 and are qualitatively identical to those reported in our previous work Merminod *et al.* (2014). In the fluid phase, for $\varepsilon \lesssim 20$, we observe at long time the classic diffusive regime, with a linear growth. The self-diffusion coefficient of particles in 2D, $D$ is determined by the slopes of the MSD, according to the relation $\mathrm{MSD}(t) \sim 4D\,t$. In the solid phase for $\varepsilon \gtrsim 200$, the MSD saturate at long time, due to the confinement of particles around the positions given by the hexagonal lattice. The maximal values of MSD decrease with $\varepsilon$ in accordance with the diminution of the kinetic energy with $\varepsilon$ Merminod *et al.* (2014); Castillo *et al.* (2020).

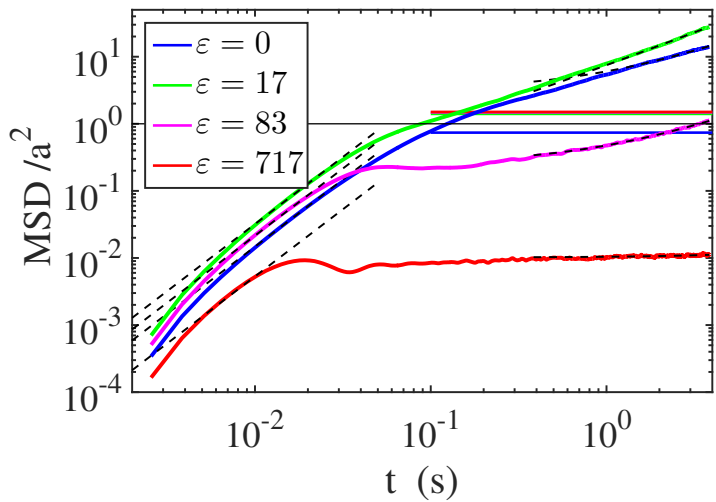

FIGURE 15. Mean square displacements $\langle |\mathbf{r}_i(t + t_0) - \mathbf{r}_i(t_0)|^2 \rangle_i$ plotted as a function of time $t - t_0$ for the four regimes defined in the main text. Displacements are normalized by the particle diameter $a$. The dashed lines are the corresponding fits, linear $4\,D\,t + \mathrm{Cste}$ at long time to characterize the diffusive behavior and $(v_{\mathrm{bal}}\,t)^2$ at short time for the ballistic regime. The thin black horizontal line corresponds to a MSD equals to $a^2$. The thick horizontal lines indicate the values of the squares of Wigner-Seitz radii $a_{WS}^2 = (\pi\,\rho_N)^{-1} = (\pi\,N/S)^{-1}$ (the lines for $\varepsilon = 17$, 83 and 717 nearly coincide). This radius corresponds to the domain available to one particle. If the MSD at long time do not exceed this limit, the particles motions are confined and do not change of neighbors, as expected for a solid phase.



\end{document}